\documentclass[11pt,preprint]{aastex}

\begin{document}

\shorttitle{LkCa 15 Disk Structure}

\shortauthors{Andrews et al.}

\title{A Closer Look at the LkCa 15 Protoplanetary Disk}

\author{Sean M. Andrews\altaffilmark{1}, Katherine A. Rosenfeld\altaffilmark{1}, David J. Wilner\altaffilmark{1}, and Michael Bremer\altaffilmark{2}}
\altaffiltext{1}{Harvard-Smithsonian Center for Astrophysics, 60 Garden Street, Cambridge, MA 02138}
\altaffiltext{2}{IRAM, 300 Rue de la Piscine, F-38406 Saint Martin d'H{\`{e}}res, France}

\begin{abstract}
We present 870\,$\mu$m observations of dust continuum emission from the LkCa 15 
protoplanetary disk at high angular resolution (with a characteristic scale of 
0\farcs25 = 35 AU), obtained with the IRAM Plateau de Bure interferometer and 
supplemented by slightly lower resolution observations from the Submillimeter 
Array.  We fit these data with simple morphological models to characterize the 
spectacular ring-like emission structure of this disk.  Our analysis indicates 
that a small amount of 870\,$\mu$m dust emission ($\sim$5\,mJy) originates 
inside a large (40-50\,AU radius) low optical depth cavity.  This result can be 
interpreted either in the context of an abrupt decrease by a factor of $\sim$5 
in the radial distribution of millimeter-sized dust grains or as indirect 
evidence for a gap in the disk, in agreement with previous inferences from the 
unresolved infrared spectrum and scattered light images.  A preliminary model 
focused on the latter possibility suggests the presence of a low-mass 
(planetary) companion, having properties commensurate with those inferred from 
the recent discovery of LkCa 15b.
\end{abstract}

\section{Introduction}

Hundreds of exoplanets have been discovered around main-sequence stars, and 
substantial effort is being invested to explain their demographics with 
formation models \citep[e.g.,][]{ida04,mordasini09}.  But associating exoplanet 
properties with their formation epoch is problematic: dramatic evolutionary 
processes that occur at early times are closely tied to the unknown physical 
conditions in the progenitor circumstellar disk.  Ideally, mature exoplanets 
could be compared with their younger counterparts, still embedded in their 
natal disks.  However, detecting planets around young stars is difficult.  
Radial velocity and transit searches are hindered by stellar variability 
\citep[e.g.,][]{huelamo08}, and direct imaging is limited by contrast with the 
bright star and disk emission.  However, the presence of a young planet can be 
inferred {\it indirectly} through its dynamical imprint on the structure of the 
disk material.  A sufficiently massive planet ($\ge 1$\,M$_{\rm Jup}$) opens a 
gap that impedes the inward flow of mass through the disk, decreasing the 
densities at the disk center \citep[e.g.,][]{lin86,bryden99,quillen04}.  The 
location of the gap marks the planet orbit, and the amount of material that 
flows across it depends on the planet mass \citep{lubow06,varniere06}.  In 
principle, the orbit and mass of a $\sim$Myr-old giant planet can be estimated
from observations of its disk birthsite, through constraints on the gap 
location and the amount of material interior to it, respectively.  

The disk around the young star LkCa 15 is considered an excellent candidate for 
planet-induced disk clearing, based on its distinctive infrared spectrum 
\citep{espaillat07} as well as the ring-like morphology of its mm-wave dust 
emission \citep{pietu06,andrews11} and scattered light in the infrared 
\citep{thalmann10}.  Those observations confirm that the LkCa 15 disk has a 
large central ``cavity", with significantly diminished dust optical depths on 
Solar System size-scales.  However, the cavity is not empty.  A faint infrared 
signal is detected in excess of the stellar photosphere, indicating that at 
least a small amount of warm dust resides near the star \citep{espaillat08}.  
That excess verifies the presence of a tenuous inner disk -- and therefore a 
gap -- although it provides only minimal bounds on its size (and therefore the 
gap width) and mass.  Based on an attempt to model a high resolution 
Submillimeter Array (SMA) observation of the LkCa 15 disk, \citet{andrews11} 
identified preliminary evidence for weak, optically thin 870\,$\mu$m emission 
from dust {\it inside} the disk cavity.  If confirmed, that emission can be 
used to estimate the inner disk mass, a key diagnostic of the flow rate across 
the gap.  

In this Letter, we present new 870\,$\mu$m continuum observations of the LkCa 
15 protoplanetary disk, with a 50\%\ improvement in angular resolution 
facilitated by the recent commissioning of high-frequency receivers at the 
Plateau de Bure interferometer (PdBI).  In \S 2, we provide a brief overview of 
the new data and describe how their combination with previous SMA observations 
provide the sharpest view yet of the thermal emission from the LkCa 15 disk.  
In \S 3 we use simple models to explore the properties of the disk cavity and 
its contents.  And in \S 4 we discuss those modeling results in the contexts of 
planet formation around LkCa 15 and the potential future utility of similar 
observations as an independent check on the properties of young exoplanets.

\section{Observations and Data Reduction}

LkCa 15 was observed for 5 hours with the most extended configuration (A: 
baselines of 130-760\,m) of the PdBI on 2011 January 27.  The observations were 
conducted in ``shared-risk" mode since they used the new Band 4 receivers at an 
effective continuum frequency of 345.8\,GHz (868\,$\mu$m) and the new WideX 
correlator to sample the continuum emission with a total bandwidth of 3.6\,GHz 
(per polarization).  The observations cycled between LkCa 15 and two nearby 
quasars, J0530+1331 and J0336+3218, every 22 minutes.  The data were calibrated 
with the {\tt CLIC} software in the {\tt GILDAS} package.  Short observations 
of the bright quasars 3C 454.3 and 3C 273 were used to set the bandpass and 
absolute flux scale, and the nearby quasars that were interleaved in the 
observing cycle were utilized to calibrate the time-dependent complex gain 
response of the system.  At the time of the observations, the new Band 4 LO 
system perturbed the first channel (of 3) in the PdBI water vapor radiometer 
(WVR) phase correction system.  We reduced the WVR system to a dual channel 
mode in the post-processing, and smoothed the WVR data on 5\,s intervals.  The 
differential phase correction determined on 45\,s intervals was extended over 
each source cycle by fitting and removing linear instrumental drifts.  This 
process requires a stable atmosphere, with water vapor fluctuations that 
average to near zero over the source cycle.  These conditions were generally 
met, due to the low water vapor levels ($<$2\,mm) present throughout the 
observations.  

To improve the Fourier coverage on short spacings, we supplemented these PdBI 
observations with the SMA data described by \citet[][baselines of 
8-508\,m]{andrews11}.  After adjusting the datasets to account for the small 
proper motion of LkCa 15 \citep{ducourant05}, the disk centroid was estimated 
in each dataset by minimizing the imaginary components of the visibilities 
\citep[see][]{andrews11}.  The inferred reference centers for the two datasets 
agree within $\sim$10\,mas and are $<$70\,mas from the expected stellar 
position (within the absolute astrometric uncertainty in each dataset), at RA = 
4$^{\rm h}$39$^{\rm m}$17\fs80 and DEC = $+$22\degr21\arcmin03\farcs20.  The 
SMA and PdBI calibrations were compared over their redundant Fourier coverage, 
and were found to be in excellent agreement on 150-500\,k$\lambda$ baselines: 
deviations between the visibility amplitudes in each dataset are random, with 
an RMS difference of $<$5\%.  The combined SMA and PdBI visibilities were 
Fourier inverted assuming natural weighting, deconvolved with the {\tt CLEAN} 
algorithm, and restored with a $0\farcs33\times0\farcs22$ synthesized beam 
using the {\tt MIRIAD} software package.  The resulting synthesized continuum 
map is shown in Figure \ref{image}, with an effective wavelength of 
870\,$\mu$m, RMS noise of 0.7\,mJy beam$^{-1}$, peak flux density of 27\,mJy 
beam$^{-1}$, and integrated flux density of 380\,mJy.

\section{Results}

\begin{figure}
\epsscale{0.65}
\plotone{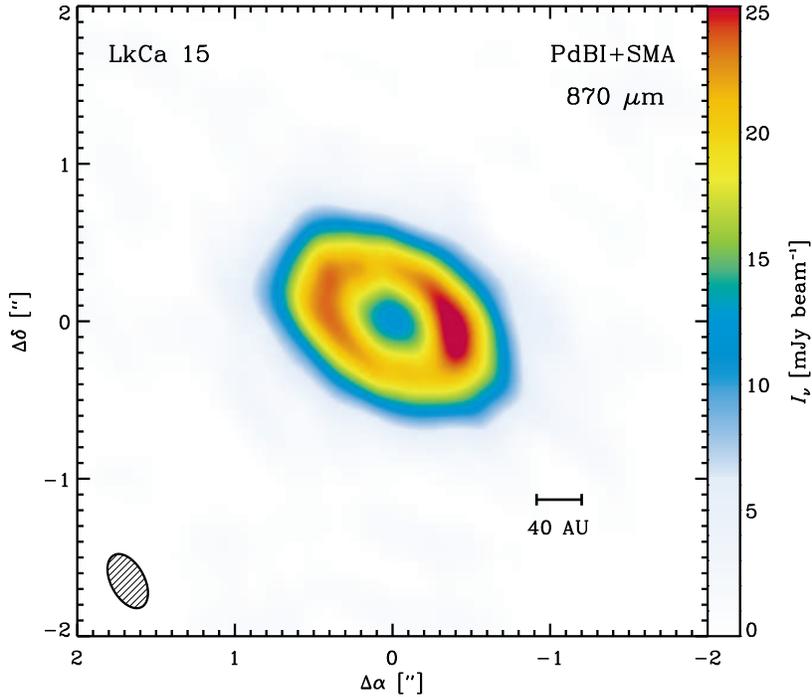}
\figcaption{Aperture synthesis image of the 870\,$\mu$m continuum emission from 
the LkCa 15 disk, made from the naturally-weighted combination of PdBI and SMA 
datasets.  The synthesized beam, with dimensions of $0\farcs33\times0\farcs22$ 
($46\times31$\,AU), is shown in the lower left.  The wedge on the right marks 
the conversion from color to surface brightness.  Each side of the image 
corresponds to 560\,AU projected on the sky.  \label{image}}
\end{figure}

\begin{figure}
\epsscale{0.55}
\plotone{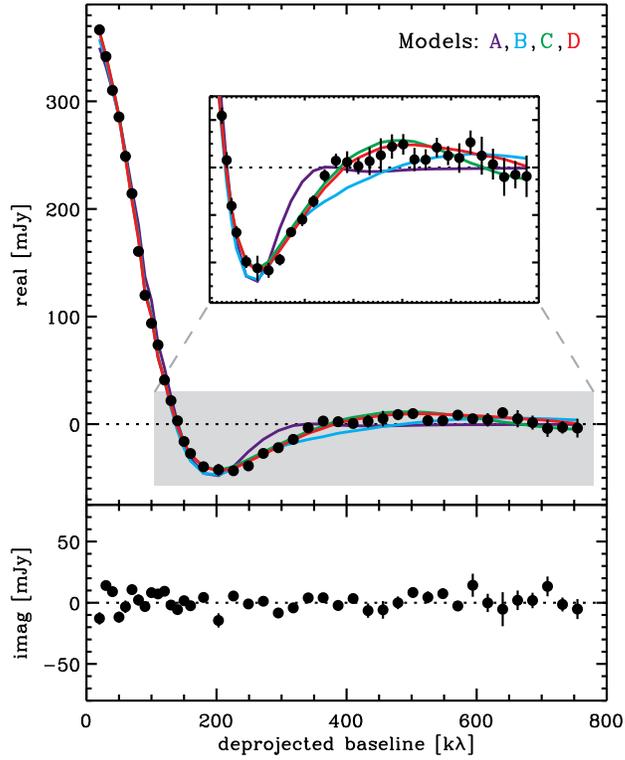}
\figcaption{The real and imaginary 870\,$\mu$m visibilities as a function of 
baseline length, deprojected to account for the LkCa 15 disk viewing geometry 
and azimuthally averaged.  The inset in the top panel is a detailed view of the 
gray-filled region.  The best-fit models visibilities for different emission 
prescriptions are overlaid in color (all models have zero imaginary fluxes, by 
definition).  \label{vis}}
\end{figure}

The 870\,$\mu$m image in Figure \ref{image} provides the sharpest view yet of 
cool dust emission from the LkCa 15 disk.  As noted previously at lower 
resolution \citep{pietu06,andrews11}, this emission has an inclined ring 
morphology with a large and prominent central depression in intensity.  The 
emission ring peaks at semimajor separations of $\sim$0\farcs4 (56\,AU for an 
assumed distance of 140\,pc) and has an aspect ratio and orientation in good 
agreement with the inclination ($i = 51\degr$) and major axis position angle 
(PA = 61\degr) inferred from its molecular line emission \citep{pietu07}.  
Figure \ref{vis} shows the azimuthally-averaged visibilities as a function of 
the deprojected baseline length (accounting for the disk viewing geometry).  
The real part of this visibility profile exhibits the classic oscillation 
pattern expected from the Fourier transform of a ring in the sky-plane, with 
distinct nulls (sign changes) at deprojected baselines near 150, 350, and 
700\,k$\lambda$.  The imaginary terms are negligible on all baselines, 
consistent with an axisymmetric emission distribution.  Although subtle, two 
qualitative features in the data can serve as useful benchmarks in a refined 
effort to characterize the LkCa 15 disk structure.  First, the continuum 
intensities inside the ring are small, but not zero (see Figure \ref{image}).  
And second, the oscillations in the continuum visibility profile are relatively 
muted, with a maximum amplitude of only $\sim$5\,mJy between the second and 
third nulls.  This latter property suggests that the emission peak near the 
inner ring edge is not very sharp.  

With those features in mind, we attempted to reproduce these LkCa 15 disk 
observations with simple 870\,$\mu$m emission models.  We adopted a radial 
surface brightness prescription that assumes optically thin thermal emission, 
$I_{\nu} \propto B_{\nu}(T_d) (1-e^{-\tau}) \approx B_{\nu}(T_d) \tau$, where 
$B_{\nu}$ is the Planck function, $T_d$ the dust temperature, and $\tau$ the 
optical depth.  The temperature profile was fixed to $T_d(R) = 100(R/{\rm 
1\,AU})^{-0.5}$\,K, based on a crude approximation of the midplane temperatures 
derived in a more sophisticated treatment of radiative transfer 
\citep{andrews11}.  Assuming the dust emissivity is independent of radius, we 
utilized a parametric form for the base optical depth profile motivated by the 
surface densities in idealized viscous accretion disks: $\tau_b(R) \propto 
(R/R_c)^{-\gamma} \exp{[-(R/R_c)^{2-\gamma}]}$ \citep[e.g.,][]{hartmann98}.  
Modifications to that base model were also considered, including an optical 
depth cavity where $\tau(R \le R_{\rm cav}) = \delta \tau_b$.  Three model 
permutations were investigated: the base model ($\delta = 1$, $R_{\rm cav}$ 
undefined; Model A), the base model with an empty cavity ($\delta = 0$; Model 
B), and the base model with a partially filled cavity ($0 < \delta < 1$; Model 
C).  All models have three base parameters -- a gradient ($\gamma$), 
characteristic size ($R_c$), and normalization (defined as the flux density, 
$F_{\rm tot} = \int I_{\nu} d\Omega$) -- and can utilize up to two additional 
parameters, \{$R_{\rm cav}$, $\delta$\}.

\begin{figure}
\epsscale{1.00}
\plotone{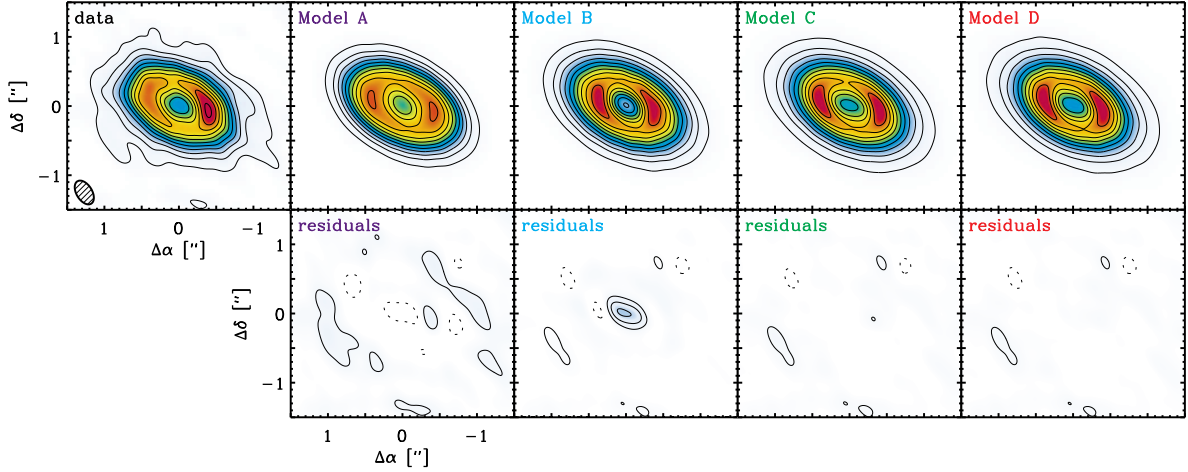}
\figcaption{Comparison of the data and models in the image plane.  The top left 
panel shows the same image as in Figure \ref{image}.  To the right, the top 
panels display the best-fit model images, and the bottom panels the imaged 
residual visibilities.  All panels show the same color scale and contour 
levels, starting at 1.4\,mJy beam$^{-1}$ (2\,$\sigma$) and increasing in 
2.5\,mJy beam$^{-1}$ (3.5\,$\sigma$) increments.  As noted in Figure \ref{vis}, 
Models C and D -- which emulate a low-density (but not empty) cavity and a gap 
structure for the LkCa 15 disk, respectively -- provide the best matches to the 
data.  \label{modims}}
\end{figure}

\begin{figure}
\epsscale{0.55}
\plotone{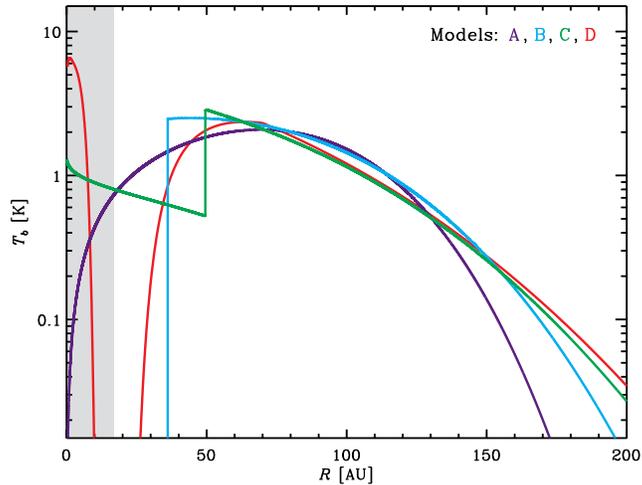}
\figcaption{Radial surface brightness profiles for the best-fit parameters of 
each model type, cast for simplicity into a brightness temperature format.  The 
combined PdBI+SMA data provide a maximum projected radial resolution of 
$\sim$17\,AU, marked here by the shaded gray region.  \label{sbs}}
\end{figure}

For a given model type and parameter set, synthetic visibilities were computed 
for the appropriate viewing geometry at the spatial frequencies observed by the 
SMA and PdBI.  Those model visibilities were compared with the data and 
assigned a fit quality statistic, the sum of the (real and imaginary) $\chi^2$ 
values over all spatial frequencies.  The best-fit parameter values for a given 
model were determined by minimizing $\chi^2$ with the Metropolis algorithm, 
utilizing multiple Monte Carlo Markov chains and an initial period of simulated 
annealing \citep[see][]{gregory05}.  The results are compiled in Table 
\ref{params_table}.  The estimated parameter uncertainties do not consider 
correlated errors from the (fixed) temperature profile or viewing geometry, and 
therefore are clearly under-estimated.  The best-fit synthetic data products 
for each model type are directly compared with the observations in Figures 
\ref{vis} and \ref{modims}.  The corresponding radial brightness profiles are 
shown together in Figure \ref{sbs}.  

For Model A, the observed emission morphology can only be reproduced with a 
large and negative optical depth gradient parameter, $\gamma$ \citep[e.g., 
see][]{isella09}.  The Model A fit does a relatively poor job accounting for 
the breadth of the observed ring structure: there is a tendency to over-predict 
the emission in the disk center and prematurely cut off at larger radii.  
Significant improvement is made with Model B, when a cavity is added to the 
base model.  This is effectively the same structure assumed by 
\citet{andrews11}.  That preliminary work used a fixed $\gamma = 1$, which 
tends to maximize the peak-to-cavity emission contrast in the fits, leading to 
higher positive residuals at the disk center.  Similar results were obtained 
when that effort was repeated here, with strong centralized residuals 
($\sim$11\,$\sigma = 8.2$\,mJy).  The best-fit Model B parameters are different 
than the fixed-gradient case (see Table \ref{params_table}; $\chi^2_{\rm A} - 
\chi^2_{\rm B} = 135$)\footnote{The $\chi^2$ differences in our progression of 
models are large enough (the best-fit likelihood ratios are significantly 
greater than unity) to warrant the complexity of adding a parameter at each 
step from Models A through C.} -- but those same residuals remain significant 
($\sim$7\,$\sigma$ = 4.8\,mJy).  Naturally, this motivated the addition of an 
emission component inside the disk cavity, cast for simplicity as an adjustment 
to $\delta$ (Model C).  The inclusion of that weak emission improved the fit 
quality ($\chi^2_{\rm B} - \chi^2_{\rm C} = 111$), leaving no significant 
residuals compared to the data.  In this scenario, dust inside the disk cavity 
produces $\sim$5\,mJy of 870\,$\mu$m emission, corresponding to 20\%\ of the 
peak surface brightness and only 1\%\ of the integrated flux density.

A gap structure represents an alternative model that naturally produces dust 
emission inside a disk cavity.  To explore that possibility with a more 
physically motivated prescription, we modeled the data with the treatment of 
gap profiles advocated by \citet{crida06} and \citet{crida07}.  In this 
scenario (Model D), we utilized a semi-analytic approximation for the surface 
density perturbation produced by an embedded low-mass companion to modify the 
base optical depth profile.  The depth, width, and shape of the gap profile 
perturbation were characterized by \citet[][their Eq.~14]{crida06} in terms of 
the companion-to-star mass ratio ($q = M_s/M_{\ast}$), the semimajor axis of 
the companion ($R_s$), the disk viscosity ($\nu$), and the local disk aspect 
ratio ($H/R$, where $H$ is the vertical scale height of the gas).  Following 
Crida and his colleagues, we fixed $H/R = 0.05$ and only investigated models 
where $\nu = 10^{-5}$ in the \citet{crida06} normalized units \citep[for our 
fixed $T_d$ profile, this corresponds to a typical viscosity coefficient 
$\alpha \sim 0.001$ in the formulation of][]{shakura73}.  Furthermore, we fixed 
$R_s = 16$\,AU, in line with the recent detection of a faint companion 
\citep[][see \S 4]{kraus11b}.  With these simplifying assumptions, Model D has 
four parameters, \{$\gamma$, $R_c$, $F_{\rm tot}$, $q$\}.  The Model D 
structure also has improved fit quality relative to the empty cavity model 
($\chi^2_{\rm B} - \chi^2_{\rm D} = 116$, comparable to Model C).  The estimate 
of $q$ implies a companion mass of $M_s = 9\pm1$\,M$_{\rm Jup}$, given the LkCa 
15 stellar mass of $M_{\ast} = 1.01 \pm 0.03$\,M$_{\odot}$ that was determined 
dynamically by \citet{pietu07}.  We should again caution that these represent 
formal parameter uncertainty estimates that are only applicable under the 
restrained assumptions of this particular model: the true uncertainties could 
be significantly larger.  As for Model C, there is roughly 5\,mJy of 
870\,$\mu$m emission interior to the gap of the favored Model D structure.

\section{Discussion}

We have used high angular resolution 870\,$\mu$m PdBI+SMA observations to 
investigate the radial distribution of cool dust in the LkCa 15 protoplanetary 
disk with simple emission models.  Although grounded in more sophisticated 
techniques, these models are inherently more morphological than physical.  
Their advantage lies in computation speed, which facilitated a broader 
exploration of dust structures that would have been prohibitive for a complex 
radiative transfer analysis.  Despite their limitations, these simple models 
provide some fundamental qualitative insights on the LkCa 15 disk properties: 
(1) there is a substantial decrease in the dust optical depths inside $R 
\approx 40$-50\,AU; (2) the emission just outside that cavity edge is not 
sharply peaked, as attested by the smooth intensity profiles produced by the 
favored negative optical depth gradients ($\gamma$); and (3) there is a small 
amount of dust located inside the disk cavity.  Given our limited resolution, 
the spatial distribution of that weak emission in the cavity is unclear.  It 
may fill the cavity (Model C), or it may be more centrally concentrated in the 
form of a gap structure (Model D) similar to what was inferred from models of 
the unresolved infrared spectrum \citep{espaillat08}.

If the latter is true, the gap is most likely opened by the resonant torques 
generated by interactions between the disk and a low-mass companion 
\citep{lin86,bryden99}.  Alternative gap-opening mechanisms -- for example, 
photoevaporation -- are unlikely given the properties of the LkCa 15 system 
\citep{alexander09,owen11}.  High-contrast imaging has ruled out stellar and 
brown dwarf companions around LkCa 15, hinting that the gap may be opened by a 
young giant planet \citep{thalmann10,pott10,kraus11}.  Recently, 
\citet{kraus11b} used a non-redundant masking technique to detect a faint, 
co-moving companion $\sim$0\farcs07 from LkCa 15.  If that object is co-planar 
with the disk and on a circular orbit, it has a semimajor axis of 16\,AU.  
Using a simple emission model based on the prescription of \citet{crida06}, we 
have shown that a gap at this location can reproduce well the resolved 
870\,$\mu$m emission morphology we observe if the companion mass is 
$\sim$9\,M$_{\rm Jup}$.  At ages of 1-3\,Myr, the \citet{baraffe03} evolution 
models suggest that this object should have an infrared contrast of $\Delta K = 
6.4$-7.2, in reasonable agreement with the $\Delta K = 6.8$ measured by 
\citet{kraus11b}.  However, the \citet{marley07} models suggest it would be 
$\sim$150$\times$ fainter: a substantial accretion luminosity would be required 
to account for the observed infrared emission.  

Ultimately, improved constraints on the companion mass could be based on the 
disk contents interior to the gap.  A crude estimate of the dust mass in that 
region can be made from the luminosity of the optically thin 870\,$\mu$m 
emission that was inferred in Models C and D.  Assuming a dust opacity of 
3\,cm$^2$ g$^{-1}$ and a fiducial $T_d = 45$\,K, the estimated flux density of 
5\,mJy corresponds to 10$^{-6}$\,M$_{\odot}$ (0.4\,M$_{\oplus}$).  If that dust 
traces the gas at a mass fraction of $\sim$1\%, then the accretion rate onto 
LkCa 15 \citep[$\dot{M}_{\ast} \approx 2 \times 10^{-9}$\,M$_{\odot}$ 
yr$^{-1}$;][]{ingleby09} implies that this inner disk material would rapidly 
drain onto the star (in $<$0.05\,Myr).  Given the system age of 1-3\,Myr, the 
inner disk must be continually replenished from the massive reservoir outside 
the gap.  There is some notable tension with theoretical expectations here: it 
is not clear how a $\sim$9\,M$_{\rm Jup}$ companion can be reconciled with the 
inferred inner disk mass and stellar accretion rate in numerical simulations of 
gap-crossing flows \citep{lubow99,lubow06}.  If LkCa 15b has a much lower mass, 
it likely cannot sculpt the deep, wide gap needed to explain the observations: 
an additional companion with a longer orbital period must also be present.  
\citet{zhu11} and \citet{dodson-robinson11} have effectively argued for this 
latter possibility.  They suggested that multi-planet systems can alleviate the 
apparent discrepancy between large transition disk cavities and accretion 
rates, implying that LkCa 15b is but one component in a young planetary system. 

Robust, quantitative constraints on the properties of LkCa 15b based on the 
structure of the LkCa 15 disk requires more work, including a stronger link 
between numerical simulations, an improved modeling effort, and observations 
that can probe the inner disk at even higher angular resolution.  Nevertheless, 
the PdBI+SMA data presented here offer a tantalizing foreshadowing of the new 
roles mm-wave observations of disk structures can play in exoplanet science.

\acknowledgments We are very grateful to Adam Kraus for his advice and for 
kindly sharing results prior to publication.  This article is based on 
observations carried out with the IRAM Plateau de Bure Interferometer and the
Submillimeter Array.  IRAM is supported by INSU/CNRS (France), MPG (Germany),
and IGN (Spain).  The SMA is a joint project between the Smithsonian 
Astrophysical Observatory and the Academia Sinica Institute of Astronomy and 
Astrophysics and is funded by the Smithsonian Institution and the Academia 
Sinica.

\clearpage

\begin{deluxetable}{lcccc}
\tablecolumns{5}
\tablewidth{0pc}
\tabletypesize{\small}
\tablecaption{Model Parameters\label{params_table}}
\tablehead{
\colhead{Model}     & \colhead{A} & \colhead{B} & \colhead{C} & \colhead{D}}
\startdata
$F_{\rm tot}$ (mJy) & $363\pm2$    & $373\pm2$    & $367\pm3$     & $385\pm2$ \\
$\gamma$            & $-1.7\pm0.1$ & $-1.0\pm0.1$ & $-0.5\pm0.1$  & $-0.3\pm0.1$ \\
$R_c$ (AU)          & $107\pm2$    & $113\pm1$    & $114\pm1$     & $113\pm1$ \\
$R_{\rm cav}$ (mJy) & \nodata      & $36\pm1$     & $49\pm1$      & \nodata   \\
$\delta$            & 1 (fixed)    & 0 (fixed)    & $0.18\pm0.02$ & \nodata   \\
$R_s$ (AU)          & \nodata      & \nodata      & \nodata       & 16 (fixed) \\
$q$                 & \nodata      & \nodata      & \nodata       & $0.009\pm0.001$ \\
\hline
$\chi^2$            & 516,735      & 516,600      & 516,489       & 516,484
\enddata
\tablecomments{Parameter estimates, formal uncertainties, and $\chi^2$ values 
for the models discussed in \S 3.  There are 776,966 independent visibilities 
used in the model fits.}
\end{deluxetable}

\clearpage

\end{document}